# Giant and supergiant electrical capacity of heterostructures on a basis of advanced superionic conductors

Alexander L. Despotuli*, Alexandra V. Andreeva*, and Peter P. Maltsev♣

*Nanoionics Laboratory, Institute of Microelectronics Technology RAS, Chernogolovka, Moscow Region, Russia

♣Department of Microsystems, Moscow State Institute of Radiotechniques, Electronics and Automatics, Moscow, Russia

Advanced superionic conductor (ASIC)/electrochemical indifferent electrode (IE) heterostructures are the key elements of capacitors with double electric layer (DEL). We show for the first time that the specific capacity ($\rho_C$) of the ASIC/smooth IE heterojunctions can be considerably exceed $\sim 10$ μFcm$^{-2}$ at the frequencies $f >> 10^{-2}$-$10^{-1}$ Hz. Heterostructures of three types with the special ASIC/smooth IE interface design (different pairs of substances) were created. They show "capacitor-like" and "battery-like" behaviors and accumulate the giant charge. It is revealed: (i) the giant capacity ($\rho_C \approx 100$ μFcm$^{-2}$ at $f \sim 2 \cdot 10^5$ Hz and $\rho_C \approx 300$ μFcm$^{-2}$ at $f \sim 10^4$ Hz, 300 K) and giant charge density $\rho_Q \approx 2(4) \cdot 10^{-4}$ C cm$^{-2}$ at 300 (370) K for the «capacitor-like» behavior; (ii) in all heterostructures the transition from "capacitor-like" behavior to "battery-like" one occurs at the critical value $\rho_{Qcr}$ ($> 2 \cdot 10^{-4}$ C cm$^{-2}$) and it allows accumulate the charge in $10^4$-$10^5$ times more; (iii) the voltages of discharge plateaus grow with increase of applied external voltage, that can not be connected with electrochemical formation of new phase with permanent composition; (iv) the supergiant $\rho_C \sim 200000$ μFcm$^{-2}$ ($f \sim 10$ Hz, 300-440K) $\rho_C \sim 10000$ μFcm$^{-2}$ ($f \sim 10^5$ Hz, 440K) were discovered at the «capacitor-like» behavior. The results of work can find applications in the area of microsources with high energy $\rho_E$ and power $\rho_W$ densities.

## 1. INTRODUCTION

The response of solid state ionic material/electrochemical indifferent electrode (IE) interfaces to external electric fields is of both fundamental and practical interest. For instance, the main obstacle for development of nano(micro)system technology (NMST) and wireless microsensors and microrobots networks (WN) is the absence of autonomous microsourses with high energy $\rho_E$ and power $\rho_W$ densities. Traditional approaches to the creation of devices for energy storage are based on rational use of volume. However, at the nano- and microscale "surface-to-volume" relation is large. Therefore, it needs the maximum use of interface properties to achieve effective energy storage and power generation.

Some types of microsources (thin-film rechargeable batteries and supercapacitors) can be made on the basis of solids with fast ion transport (FIT), i.e. on the basis of solid electrolytes and superionic conductors (SICs) [1]. Advanced superionic conductors (ASICs) present the subclass of SICs [2-4]. ASICs are solids with specific crystal structure (close to optimal for FIT), record high level of ionic conductivity $\sigma_i$ ($>0.1$ Ohm$^{-1}$cm$^{-1}$ 300 K) and small activation energy of FIT ($\approx 0.1$ eV). Decades, ASICs are used in supercapacitors - devices where the energy and charge store in double electric layers (DELs) at the ASIC/IE functional interfaces. The DEL-thickness is an order of molecule size. In the existing designs of supercapacitors with liquid electrolytes and volume-distributed electrodes from various kinds of nanostructured carbon, specific capacities are $\approx 100$ F/g at internal surfaces $\approx 10^7$ cm$^2$/g [5], which gives a true $\rho_C$ value of $\sim 10$ μF/cm$^2$. Values of the same order are characteristic of DELs on ASIC/electrode heterojunctions [6]. However, liquid electrolytes are not applicable in microelectronics. Solid state supercapacitors on the basis of ASIC have DEL with real surface capacity $\rho_C \approx 100$ μF/cm$^2$ only at low frequencies ($f \sim 10^{-2}$-$10^{-1}$ Hz) and $\rho_C \sim 10$ μF/cm$^2$ when $f >> 10^{-2}$-$10^{-1}$ Hz [6]. It looks like a paradox because the mobile ions in the ASIC crystal structure oscillate between neighbor next crystallographic positions with frequencies of $\sim 10^{10}$ Hz at 300 K. The absence of FIT in the area of ASIC/IE functional heterojunctions (in the area of DEL) is a cause of the low operational frequencies. Distortion and disorder of specific ASIC crystal structure at the heterointerfaces lead the delayed ion transport.

The microsources with different "energy-power" relation are needed for NMST and WN. Fuel cell provide $\rho_E \approx 6$ times larger but $\rho_W$ lower than those of the Li-ion microbatteries [7-9]. On the Ragone plot («$\rho_E$ - $\rho_W$») [8]. ASIC based supercapacitors occupy an intermediate place between Li-ion batteries and conventional capacitors. Supercapacitors have more $\rho_E$ (smaller $\rho_W$) than conventional capacitors. The delayed ion transport in DEL is the cause of more low $\rho_W$. The experiments on the creation of ASIC/IE lattice-matched heterojunction with FIT in DEL were first carried out in the IMT RAS (1991-1992) [4]. The appropriate results were first published in [10, 11, 13]. Recently, for the increase of $\rho_E$, $\rho_W$ and $\rho_C$ densities in DEL the conception of ASIC/IE functional coherent interfaces (with $\rho_C \sim 100$ μF/cm$^2$ at the high frequencies) was proposed [10-14].

In this work the approach «from advanced materials to advanced devices» [13] and nanoionics [15] are used to create solid-state nanoionic sources (NS) [2-4]. The aim is to create innovative heterostructures with a special interface design of heterojunctions on the ASIC basis, providing record high values of capacity and functioning frequency. The heterostructures considered are the basis of high-frequency NSs necessary for development of NMST and WNs.

## 2. EXPERIMENTAL RESULTS

Experiments on the creation of ASIC/IE coherent heterojunctions at the UHV conditions were carried out at IMT RAS in 2004-2005 for the first time. They give the patent important information. The NS laboratory samples with a special ASIC/IE heterojunction design and giant values of capacity density ($\rho_C$) and charge density ($\rho_Q$) were created. In this paper we report the electric characteristics of three samples (A, B, C) with the same type of the ASIC/IE interface design, and, simultaneously with different compositions and crystal structures of ASICs and/or IEs. The electrical characteristics of the samples were investigated by means of impulse technique. П-impulses of voltage from generator were applied to series circuit of the investigated heterostructure and load resistance. The dependences of voltage on the heterostructures during of charge-discharge process were registered by an oscilloscope.

The oscillograms of charge (duration 0.17 ms) – discharge (in coordinates "voltage-time") of the experimental heterostructure (sample A) with a smooth IE (without micro-roughness) are presented in Fig.1. On the time intervals $\approx 0.2$ ms and current densities $j \sim 0.02$ -0.3 A/cm$^2$ (300 K) the heterostructure demonstrates "capacitor-like" behavior with $\rho_C \approx 300$ μF/cm$^2$. This value found by the comparison with a charge-discharge curve (oscillogram 5) of conventional capacitor with the known capacity. At the accumulation charge density $\rho_Q \approx 2 \cdot 10^{-4}$ C/cm$^2$ (oscillogram 2), the transition to "battery-like behavior occurs (oscillogram 3). The distinctive features of the "battery-like" behavior are the flat sections

(plateaus) on the charge and discharge curves. In this mode the released charge 20 times greater (at a discharge current $j_p \approx 8$ Acm$^{-2}$) than that of oscillogram 1. So, an equivalent capacitor, corresponding to heterostructure, should have specific capacity $\rho_{C\,eq} \approx 6000$ µF cm$^{-2}$.

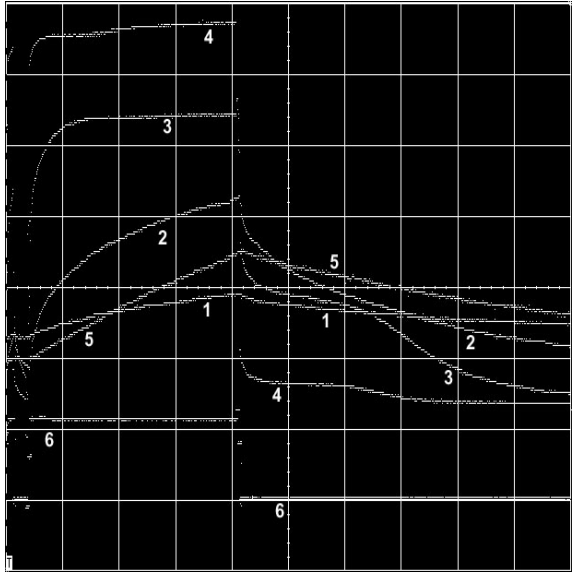

**Fig.1.** Charge-discharge oscillograms of the ASIC/smooth IE experimental heterostructure (sample A) at 300 K in coordinates "voltage–time". Horizontal scale is 0.05 ms/div.
1 - heterostructure, charge-discharge through relative resistance $100r$, discharge current density $j \approx 0,02$ A/cm$^2$, "capacitor-like" behavior, $\rho_C \approx 300$ µF/cm, $\rho_Q \approx 3\cdot10^{-5}$ C/cm$^2$; 2 - heterostructure ($10r$, $j \approx 0,3$ A/cm$^2$, "capacitor-like" behavior, $\rho_Q \approx 2\cdot10^{-4}$ C/cm$^2$);
3 - heterostructure ($1r$, $j \approx 2$ A/cm$^2$, transition from "hybrid" to "battery-like" behavior, $\rho_Q \approx 5\cdot10^{-4}$ C/cm$^2$); 4 - heterostructure ($0,1r$, $j_p \approx 8$ A/cm$^2$, "battery-like" behavior, failure of the charge accumulation mode); 5 - conventional capacitor (reference capacitor), charge-discharge through relative resistance $100r$; 6 – the form of impulse of applied external voltage to heterostructure.

Fig.2 shows the charge (1.7 ms) – discharge oscillograms for experimental heterostructure (A) at 300 K. The transition to "battery-like" behavior occurs within $\approx 0.5$ ms when the stored charge density $\rho_Q \approx 1.5\cdot10^{-4}$ C/cm$^2$ (oscillogram 2) is achieved. The data of Fig.1 and Fig.2 suggest the existence of the critical value $\rho_{Qcr} \sim 10^{-4}$ C/cm$^2$ (300 K). It is of the order of $\rho_Q$ on the densely packed planes (ions of the same sign) in the ionic crystals. It can be assumed that the transition to "battery-like" behavior occurs when mobile ions leave the contact layer of ASIC (nearest to IE). It also follows from Fig.2 (oscillogram 1 and 4), that $\rho_{Q4} \approx 170 \rho_{Q1}$ ($\rho_C \approx 300$ µF/cm$^2$). So, the equivalent capacitor should have $\rho_{Ceq} \approx 50000$ µF/cm$^2$ for the time intervals $\sim 2$ ms ($f \sim 1$ kHz).

The oscillograms of charge (18 ms) – discharge of the experimental heterostructure (300 K, sample A) are given in Fig.3. The transition to "battery-like" behavior occurs at $\rho_{kp} \approx 2\cdot10^{-4}$ C cm$^{-2}$, which corresponds to the data of Fig. 1 and Fig. 2. For the oscillograms 1 (Fig. 1) and 4 (Fig. 3), the values of $\rho_Q$ differ by 2000 times, so within 20 ms time intervals (frequencies $\approx 100$ Hz) the equivalent capacitor should have $\rho_{eq} \approx 600000$ µF cm$^{-2}$. The same value was obtained by the comparison with an accumulative charge of the conventional capacitor (Fig. 3, the oscillogram 5).

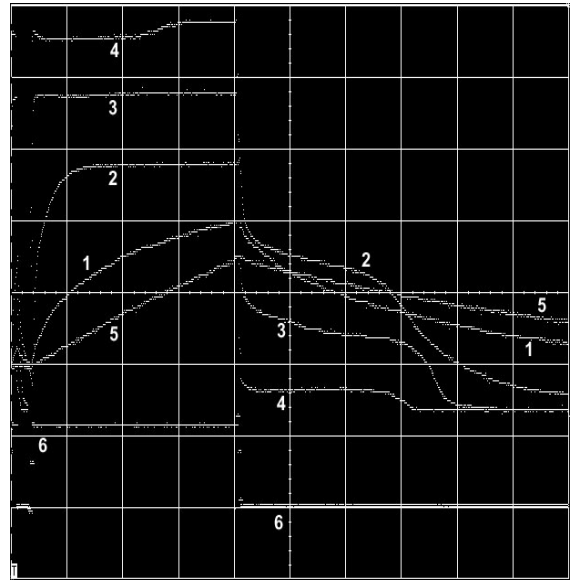

**Fig.2.** Charge-discharge oscillograms of the ASIC/smooth IE experimental heterostructure (sample A) at 300 K Horizontal scale is 0.5 ms/div.
1 - heterostructure, charge-discharge through relative resistance $100r$, discharge current $j \approx 0,03$ A/cm$^2$, "capacitor-like" behavior, $\rho_C \approx 300$ µF/cm$^2$ and $\rho_Q \approx 6\cdot10^{-5}$ C/cm$^2$ ; 2 - heterostructure ($10r$, $j \approx 0,3$ A/cm$^2$, "battery-like" behavior, $\rho_Q \approx 5\cdot10^{-4}$ C/cm$^2$); 3 - heterostructure ($1r$, $j \approx 2$ A/cm$^2$, "battery-like" behavior, $\rho_Q \approx 3\cdot10^{-3}$ C/cm$^2$); 4 - heterostructure ($0,1r$, $j \approx 8$ A/cm$^2$, "battery-like" behavior, failure of the charge accumulation mode, $\rho_Q \approx 1\cdot10^{-2}$ C/cm$^2$ ); 5 - conventional capacitor, charge-discharge through $100r$; 6 - the form of applied external voltage impulse.

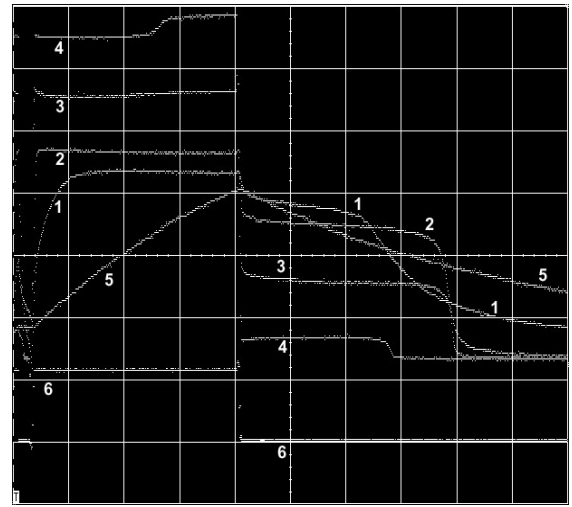

**Fig.3.** Charge-discharge oscillograms of the ASIC/smooth IE experimental heterostructure (sample A) at 300 K Horizontal scale is 5 ms/div.
1 - heterostructure, charge-discharge through relative resistance $100r$, discharge current $j \approx 0,03$ A/cm$^2$, transition to "battery-like" behavior, $\rho_Q \approx 6\cdot10^{-4}$ C/cm$^2$; 2 - heterostructure ($10r$, $j \approx 0,3$ A/cm$^2$, "battery-like" behavior, $\rho_Q \approx 5\cdot10^{-3}$C/cm$^2$); 3 - heterostructure ($1r$, $j \approx 2$ A/cm$^2$, "battery-like" behavior, $\rho_Q \approx 4\cdot10^{-2}$ C/cm$^2$); 4 - heterostructure ($0,1r$, $j \approx 5$ A/cm$^2$, "battery-like" behavior, failure of the charge accumulation mode, $\rho_Q \approx 6\cdot10^{-2}$ C/cm$^2$ ); 5 - conventional capacitor, charge-discharge through $100r$; 6 - the form of applied external voltage impulse.

Fig. 4 shows the charge (370 ms) – discharge oscillograms for experimental heterostructure (A) at 300 K. The comparison of the charge accumulation by the heterostructure within 370 ms time and by conventional capacitor (Fig.4, oscillograms 3 and 5) gives $\rho_{Ceq} \approx 600000$ μF/cm$^2$.

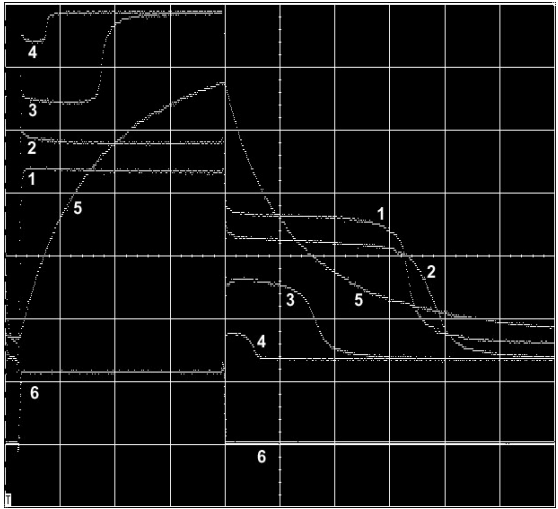

**Fig.4. Charge-discharge oscillograms of the ASIC/smooth IE experimental heterostructure (sample A) at 300 K  Horizontal scale is 100 ms/div.**
1 - heterostructure, charge-discharge through relative resistance 100$r$, discharge current $j \approx 0,03$ A/cm$^2$, "battery-like" behavior, $\rho_Q \approx 1 \cdot 10^{-2}$ C/cm$^2$; 2 - heterostructure (10$r$, $j \approx 0,3$ A/cm$^2$, "battery-like" behavior e, $\rho_Q \approx 1 \cdot 10^{-1}$C/cm$^2$); 3 - heterostructure (1$r$, $j \approx 2$ A/cm$^2$, "battery-like" behavior, failure of the charge accumulation mode, $\rho_Q \approx 3 \cdot 10^{-1}$ C/cm$^2$); 4 - heterostructure (0,1$r$, $j \approx 6$ A/cm$^2$, "battery-like" behavior, failure of the charge accumulation mode, $\rho_Q \approx 2.4 \cdot 10^{-1}$ C/cm$^2$ ); 5 - conventional capacitor, charge-discharge discharge through 100$r$; 6 - the form of applied external voltage impulse.

The oscillograms of charge ( 20 μs) – discharge of the experimental heterostructure ( A) at 300 K are given in Fig.5. The comparison of the charge accumulated by the heterostructure and by conventional capacitor (Fig.5, oscillograms 1 and 4) gives $\rho_C \approx 100$ μF cm$^{-2}$. ($f \sim 10^5$ Hz).

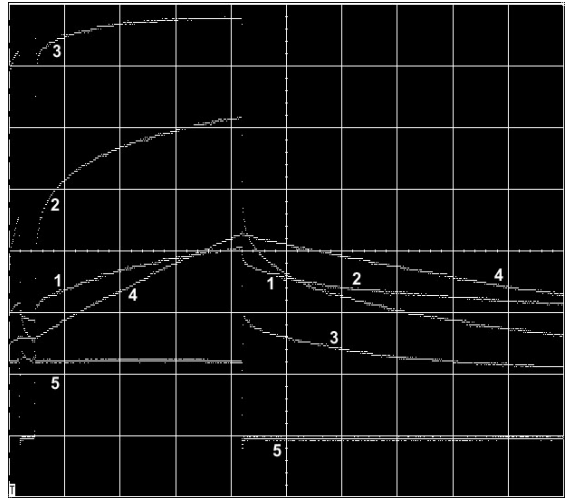

**Fig.5. Charge-discharge oscillograms of the ASIC/smooth IE experimental heterostructure (sample A) at 300 K. Horizontal scale is 5 μs/div.**
1 - heterostructure, charge-discharge through relative resistance 10$r$, discharge current $j \sim 0,2$ A/cm$^2$, "capacitor-like" behavior, $\rho_C \approx 100$ μF/cm$^2$; 2 - heterostructure (1$r$, $j \sim 0,3$ A/cm$^2$, "capacitor–like" behavior); 3 - heterostructure (0.1$r$, $j \approx 10$ A/cm$^2$, "capacitor–like" behavior); 4 - conventional capacitor, charge-discharge through 10$r$; 5 - the form of applied external voltage impulse.

The influence of temperature on the internal resistance, $\rho_C$, $\rho_{Cэк}$ and $\rho_Q$ in the experimental heterostructures was investigated. The oscillograms of charge (370 ms) – discharge of the sample A at 370 K are given in Fig.6.

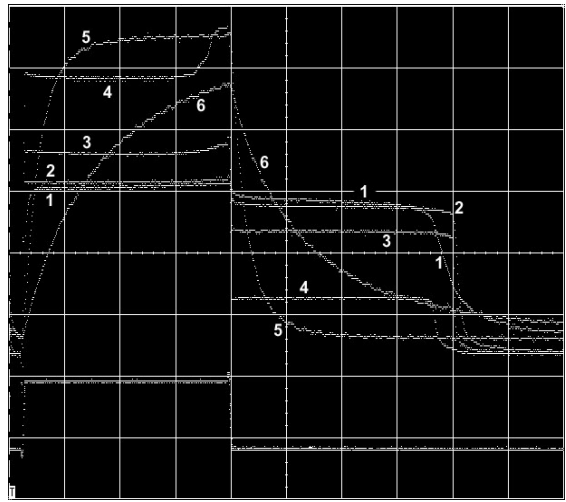

**Fig.6. Charge-discharge oscillograms of the ASIC/smooth IE experimental heterostructure (sample A) at 370 K. Horizontal scale is 100 ms/div.**
1- heterostructure, charge-discharge through relative resistance 100$r$, discharge current $j \approx 0,04$ A/cm$^2$, "battery-like" behavior, $\rho_Q \approx 1,6 \cdot 10^{-2}$ C/cm$^2$; 2 - heterostructure (10$r$, $j \approx 0,4$ A/cm$^2$,"battery-like" behavior behavior, $\rho_Q \approx 1,6 \cdot 10^{-1}$C/cm$^2$); 3 – heterostructure (1$r$, $j \approx 3$ A/cm$^2$, "battery-like" behavior, the beginning of failure of the charge accumulation mode, $\rho_Q \approx 1.2$ C/cm$^2$); 4 - heterostructure (0,1$r$, $j \approx 17$ A/cm$^2$, "battery-like" behavior, failure of the charge accumulation mode, $\rho_Q \approx 6$ C/cm$^2$ ); 5 - conventional capacitor $C_1$, charge-discharge through 100$r$ resistor; 6 - conventional capacitor $C_2=4.3 \cdot C_1$ charge-discharge through 100$r$; 7 - the form of applied external voltage impulse.

Fig.7 shows the charge (200 ms) – discharge oscillograms of the experimental heterostructure (sample B) with smooth IE (at 440 K). The heterostructure displays a distinct "capacitor-like" behavior with $\rho_C$ >200000 µF/cm$^2$ and φ ≈88° (oscillogram 1). The conventional capacitor (oscillogram 5) had φ ≈79° under the same conditions. The transition to the "battery" behavior occurs at $\rho_Q$ ~0.1 C/cm$^2$ (oscillogram 2).

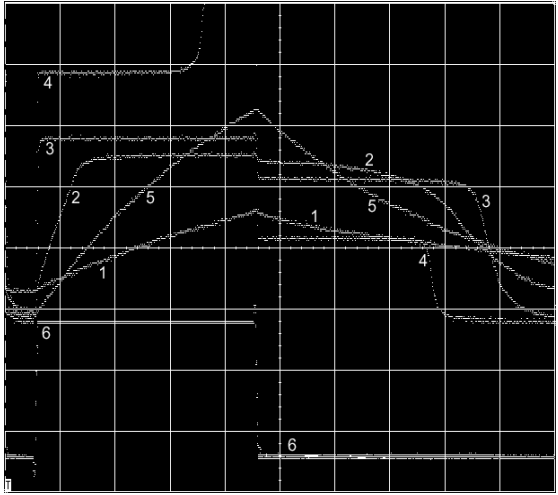

**Fig.7. Charge-discharge oscillograms of the ASIC/smooth IE experimental heterostructure (sample B) at 440 K. Horizontal scale is 50 ms/div.**
1 - heterostructure, charge-discharge through relative resistance 10$r$ relative number, discharge current $j$ ≈0.3 A/cm$^2$, "capacitor-like" behavior, $\rho_C$ >200000 µ F/cm$^2$, $\rho_Q$ ≈6·10$^{-2}$ C/cm$^2$; 2 - heterostructure (1$r$, $j$ ≈5.5 A/cm$^2$, transition to "battery-like" behavior, $\rho_Q$ ≈1 C/cm$^2$, $\rho_{C\,eq}$ ≈3.7 F/cm$^2$); 3 – heterostructure (0.1$r$, $j$ ≈50 A/cm$^2$, "battery-like" behavior, $\rho_Q$ ≈10 C/cm$^2$, $\rho_{C\,eq}$ ≈33 F/cm$^2$); 4 - heterostructure (0.01$r$, $j$ ≈260 A/cm$^2$, "battery-like" behavior failure of the charge accumulation mode, $\rho_Q$ ≈40 C/cm$^2$, $\rho_{C\,eq}$ ≈40 F/cm$^2$);
5 - conventional capacitor, charge-discharge through 10$r$;
6 - the form of applied external voltage impulse.

The oscillograms of charge (200ms) – discharge of the sample B at 300K are given in Fig.8. At the time intervals ≈200 ms and $j$~0,3-5 A/cm$^2$ the value $\rho_C$>200000 µF/cm$^2$ is achieved. Thus, the obtained experimental data of $\rho_C$ for ASIC/IE heterojunctions with special interface design considerably exceed the known experimental values [6,16] and recent theoretical estimations [3,4].

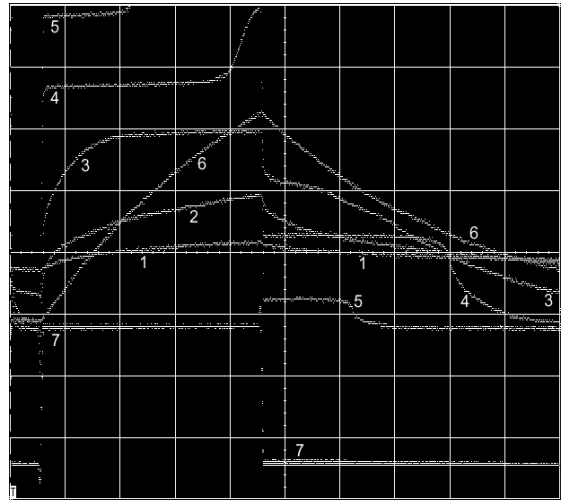

**Fig.8. Charge-discharge oscillograms of the ASIC/smooth IE experimental heterostructure (sample B) at 300 K. Horizontal scale is 50 ms/div.**
1 - heterostructure, charge-discharge through relative resistance 100$r$, discharge current $j$ ≈0.01A/cm$^2$, "capacitor-like" behavior, $\rho_C$ > 200000 µF/cm$^2$; 2 - heterostructure (10$r$, $j$ ≈0.2 A/cm$^2$, "capacitor-like" behavior, $\rho_C$ >200000 µF/cm$^2$); 3 - heterostructure (1$r$, $j$ ≈4.4 A/cm$^2$, transition to "battery-like" behavior, $\rho_Q$ ≈0.7C/cm$^2$, $\rho_{C\,eq}$ ≈2 F/cm$^2$); 4 - heterostructure (0.1$r$, $j$ ≈33 A/cm$^2$, "battery-like" behavior, beginning of the failure of the charge accumulation mode, $\rho_Q$ ≈5 C/cm$^2$, $\rho_{C\,eq}$ ≈15 F/cm$^2$); 5 - heterostructure (0.01$r$, $j$ ≈90 A/cm$^2$, "battery-like" behavior, failure of the charge accumulation mode);
6 - conventional capacitor, charge-discharge through 10$r$;
7 - the form of applied external voltage impulse.

The influence of the Π-impulse amplitude of applied external voltage on the plateau position at "battery- like" behavior was investigated. Figures 9, 10 and 11 present the charge-discharge oscillograms obtained at the same sensitivity (V/div) when the voltage steps of the fixed form were applied to the heterostructure (sample B). The ratio of the maximum impulse amplitudes in the figures is 1:2:4. Figure 9 shows that an increase of the external voltage amplitude by 2 times after the heterostructure transition to "battery-like" behavior does not affect the position of charge (discharge) plateau. However, if this transition occurs at the voltage twice as large, it is the voltage that determines the position of charge (discharge) plateaus. The charge plateau voltage virtually coincides with that of discharge. Similar conclusions can be drawn from Fig. 11, which shows that an increase in the external voltage amplitude of the step by 2 times (as compared to the corresponding step in Fig.10) shifts the charge (discharge) plateau to the position corresponding to external voltage twice as large.

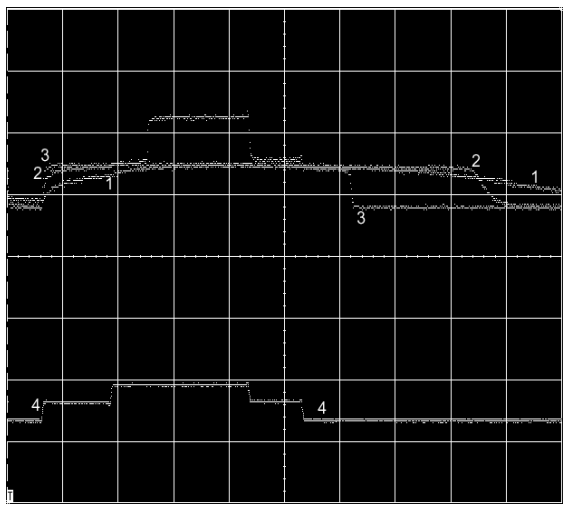

**Fig.9. Charge-discharge oscillograms of the ASIC/smooth IE experimental heterostructure (sample B) at 440 K. Horizontal scale is 100 ms/div.**
1 - heterostructure, charge-discharge through relative resistance $100r$;
2 - heterostructure, charge-discharge through $10r$;
3 – heterostructure charge-discharge through $1r$, failure of the charge accumulation mode); 4 - the form of applied external voltage impulse.

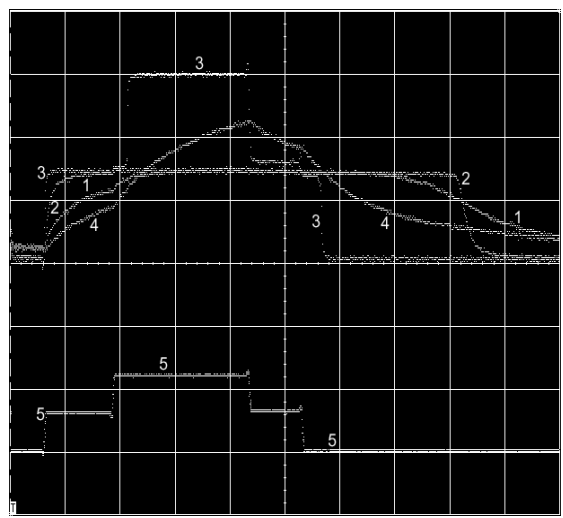

**Fig.10. Charge-discharge oscillograms of the ASIC/smooth IE experimental heterostructure (sample B) at 440 K. Horizontal scale is 100 ms/div.**
1 - heterostructure, charge-discharge through relative resistance $100r$;
2 - heterostructure, charge-discharge through $10r$;
3 – heterostructure charge-discharge through $1r$, failure of the charge accumulation mode); 4 - conventional capacitor, $100r$;
5 - the form of applied external voltage impulse.

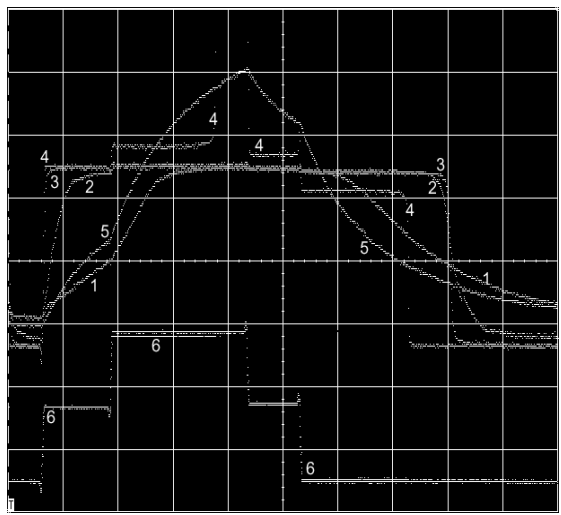

**Fig.11. Charge-discharge oscillograms of the ASIC/smooth IE experimental heterostructure (sample B) at 440 K. Horizontal scale is 100 ms/div.**
1 - heterostructure, charge-discharge through relative resistance $100r$;
2 - heterostructure, charge-discharge through $10r$;
3 - heterostructure, charge-discharge through $1r$; 4 - heterostructure charge-discharge through $0.1r$, failure of the charge accumulation mode); 5 - conventional capacitor, $100r$; 6 - the form of applied external voltage impulse.

The data of figures 9, 10 and 11 prove that no electrochemical deposition of a new phase with permanent composition occurs in the case of "battery-like" behavior. Apart from samples (A) and (B) one more ASIC/IE heterostructure (sample C) was investigated. It exhibits the same features of behavior (but with worse characteristics) as the sample (A). Fig. 12 shows the "battery-like" behavior and failure of the charge accumulation mode in the sample (C).

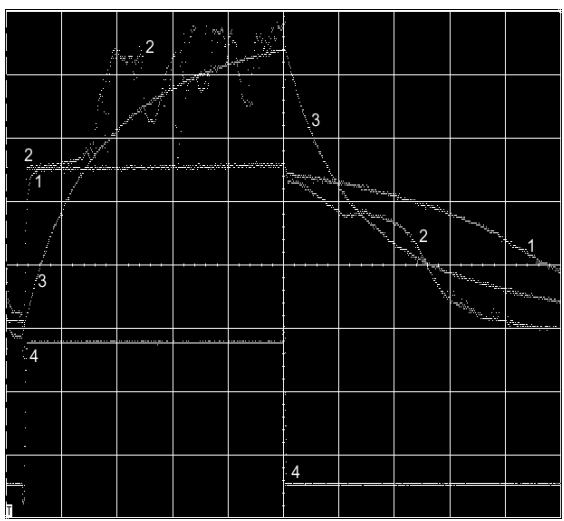

**Fig.12. Charge-discharge oscillograms of the ASIC/smooth IE experimental heterostructure (sample C) at 430 K. Horizontal scale is 100 ms/div.**
1 - heterostructure, charge-discharge through relative resistance $100r$, discharge current $j \approx 0.03$ A/cm$^2$, "battery-like" behavior, $\rho_{C\,eq} > 8000$ μF/cm$^2$; 2 - heterostructure ($10r$, $j \approx 0.2$ A/cm$^2$, failure of the charge accumulation mode, $\rho_Q \approx 6 \cdot 10^{-3}$ C/cm$^2$); 3 - conventional capacitor, charge-discharge through $100r$; 4 - the form of applied external voltage impulse.

The similarity of behavior and properties of A, B and C heterostructures having the same ASIC/IE heterojunctions design but differing in chemical composition of ASIC and/or IE structures strongly suggest the universal character of "battery-like" behavior. The following mechanism of charge accumulation can be proposed: (1) accumulation of a threshold charge density $\rho_{Q\text{crit}}$ on the ASIC/IE interface; (2) electron transfer from the ASIC valence band to the anode in strong electric field (order of molecular field) and simultaneous escape of positively charged mobile ions to the cathode, to produce neutral complex point defects on the basis of holes and cation vacancies, and (3) distribution of the neutral defect zone with the defect concentration determined by an applied external voltage) into the depth of ASIC volume via self-organization. In the case of "battery-like" behavior defect concentrations in the ASIC subsurface layers can be detected by optical methods.

The experimental observation of supergiant capacity phenomenon on the investigated heterostructures with special interface design can be connected with formation of the ordered single-layer new phase containing alternating IE-cations and ASIC-anions at the interface. The great $\rho_Q$ value may be achieved by fractal surface geometry.

## 3. CONCLUSION

The discovery of giant and supergiant capacitor phenomena as well as "battery-like" behavior in the special designed ASIC/IE heterojunctions is a fact of a great practical interest. Is there new physics on horizon?

* Electronic address: despot@ipmt-hpm.ac.ru